\documentclass[aps,preprint]{revtex4}%
\usepackage{amsfonts}
\usepackage{amsmath}
\usepackage{amssymb}
\usepackage{graphicx}%
\providecommand{\U}[1]{\protect\rule{.1in}{.1in}}

\begin{document}
\preprint{ }
\title{Massive Gravity Simplified: A Quadratic Action\\ }
\author{Ali H. Chamseddine}
\affiliation{American University of Beirut, Physics Department, Beirut, Lebanon}
\affiliation{and I.H.E.S. F-91440 Bures-sur-Yvette, France}
\affiliation{and Universit\'{e} de Tours, Parc de Grandmont, 37200 Tours, France}
\affiliation{LE STUDIUM, Loire Valley Institute for Advanced Studies, Tours and Orleans, France}
\author{Viatcheslav Mukhanov}
\affiliation{Theoretical Physics, Ludwig Maxmillians University,Theresienstr. 37, 80333
Munich, Germany and Department of Physics, New York University, NY 10003, USA}
\pacs{PACS number}

\begin{abstract}
We present a simplified formulation of massive gravity where the Higgs fields
have quadratic kinetic term. This new formulation allows us to prove in a very
explicit way that all massive gravity theories considered so far inevitably
have Boulware-Deser ghost in non-trivial fluctuations of background metric.

\end{abstract}
\maketitle

In a series of papers we have used the Higgs mechanism to give mass to the
graviton \cite{CM}, \cite{CM1}, \cite{CM2}. The requirement of Poincare
invariance imposes rather severe conditions on the possible Higgs fields.
Namely, in diffeomorphism invariant theories we are constrained to use four
scalar fields $\phi^{A},$ $A=0,1,2,3$ to play the role of Higgs fields. All
proposed Poincare and diffeomorphism invariant theories of massive gravity
(see refs. in recent review \cite{review}) can be reduced to Higgs gravity
where the massive term is built out of the following diffeomorphism invariant
combinations of the scalar fields \cite{CM}
\begin{equation}
\bar{h}^{AB}=g^{\mu\nu}\partial_{\mu}\phi^{A}\partial_{\nu}\phi^{B}-\eta^{AB}.
\label{1}%
\end{equation}
When the scalar fields acquire vacuum expectation values, proportional to the
space-time coordinates%
\begin{equation}
\left\langle \phi^{A}\right\rangle =x^{A}, \label{2}%
\end{equation}
their perturbations%
\[
\chi^{A}=\phi^{A}-x^{A},
\]
around this broken symmetry background induce extra massive metric degrees of
freedom. When combined with two degrees of freedom of GR graviton (which also
becomes massive due to interactions with the background) they constitute a
massive graviton. A Poincare invariant massive graviton must have five degrees
of freedom. However, four scalar fields in general have four degrees of
freedom, one of which, namely $\chi^{0},$ inevitably describes a ghost. There
is a unique choice of the action for scalar fields, to order $\bar{h}^{2}$,
where the ghost does not propagate around Minkowski background. This is the
well-known Fierz-Pauli action \cite{pauli}%
\begin{equation}
S_{\phi}=\frac{m_{g}^{2}}{8}\int d^{4}x\,\sqrt{-g}\left[  \bar{h}^{2}-\bar
{h}_{\,\,\,B}^{A}\bar{h}_{\,\,A}^{B}+O\left(  \bar{h}^{3}\right)  \right]  ,
\label{3}%
\end{equation}
where we raise and lower indices with Minkowski metric $\eta^{AB}.$ If we
restrict ourselves to quadratic terms and substitute in this action
\begin{equation}
\bar{h}^{AB}=h^{AB}+\partial^{A}\chi^{B}+\partial^{B}\chi^{A}+\partial^{C}%
\chi^{A}\partial_{C}\chi^{B}+h^{\mu\nu}\partial_{\mu}\chi^{A}\partial_{\nu
}\chi^{B} \label{4}%
\end{equation}
then we immediately find that up to second order in $h^{2}$, $h\chi$ and
$\chi^{2}$ the field $\chi^{0}$ is the Lagrange multiplier and hence has no
propagator. However, in the next order there appears the term $h\left(
\dot{\chi}^{0}\right)  ^{2}$ and the field $\chi^{0}$ starts to propagate on a
background which deviates from Minkowski background. This is the well known
nonlinear Boulware-Deser ghost \cite{boul}. Because one can always add to the
Fierz-Pauli term the higher order in $\bar{h}$ terms, there was a hope that
there is a unique (up to the total derivatives) action, which can be written
as an infinite series in powers of $\bar{h},$ where BD ghost does not
appear\cite{GR1}. In \cite{CM2} we have shown that even in this theory the
nonlinear ghost arises, but only in the fourth order of perturbation theory.
However, in \cite{GR1}, \cite{HR} it was claimed that using fields
redefinitions this ghost could be moved to the next orders and eventually
might be canceled.

In this paper we will show that contrary to these claims the nonlinear ghost
is inevitable in any theory described by action $\left(  \ref{4}\right)  $
irrespective of its nonlinear extension. With this purpose we first present a
new simplified reformulation of massive gravity, where the action proposed in
\cite{GR1}, will depend only quadratically on $\phi^{A}.$

Let us introduce the four vector fields $e_{A}^{\mu},$ which are constrained
to satisfy
\begin{equation}
g^{\mu\nu}=e_{A}^{\mu}e_{B}^{\nu}\eta^{AB}.\label{5}%
\end{equation}
Next we define
\begin{equation}
S_{AB}=e_{A}^{\mu}\partial_{\mu}\phi_{B}-\eta_{AB},\label{6}%
\end{equation}
which is constrained to be symmetric
\begin{equation}
S_{AB}=S_{BA},\label{7}%
\end{equation}
implying
\begin{equation}
e_{A}^{\mu}\partial_{\mu}\phi_{B}-e_{B}^{\mu}\partial_{\mu}\phi_{A}%
=0.\label{8}%
\end{equation}
There are $16$ constraints on the $16$ fields $e_{A}^{\mu}$ which could be
solved unambiguously to determine $e_{A}^{\mu}$ in terms of $g^{\mu\nu}$ and
$\partial_{\mu}\phi_{A}.$ This can only be done perturbatively, which in turn
implies that $S_{AB}$ depends on $g^{\mu\nu}$ and $\partial_{\mu}\phi_{A}$
nonlinearly$.$ Let us consider the following diffeomorphism invariant action
\begin{equation}
S=-\frac{1}{2}\int d^{4}x\,\sqrt{-g}R+\frac{m_{g}^{2}}{8}\int d^{4}%
x\,\sqrt{-g}\left[  S^{2}-S_{AB}S^{AB}+O\left(  S^{3}\right)  \right]
,\label{9}%
\end{equation}
where
\begin{equation}
S=S_{AB}\eta^{AB}.\label{10}%
\end{equation}
The constraints $\left(  \ref{5}\right)  $ and $\left(  \ref{6}\right)  $ can
be imposed by using the method of Lagrange multipliers%
\begin{equation}
\frac{1}{2}\int d^{4}x\,\sqrt{-g}\left(  \tau_{\mu\nu}\left(  g^{\mu\nu}%
-e_{A}^{\mu}e_{B}^{\nu}\eta^{AB}\right)  +2\lambda^{AB}S_{AB}\right)
,\ \label{11}%
\end{equation}
with
\begin{equation}
\tau_{\mu\nu}=\tau_{\nu\mu},\qquad\lambda^{AB}=-\lambda^{BA}\label{12}%
\end{equation}
We now show that the class of actions $\left(  \ref{9}\right)  $ is equivalent
to the class of actions $\left(  \ref{3}\right)  .$ Consider the matrices%
\begin{equation}
S_{AB}^{^{\prime}}=e_{A}^{\mu}\partial_{\mu}\phi_{B}=S_{AB}+\eta
_{AB},\label{13}%
\end{equation}
and their product
\begin{align}
S_{AC}^{^{\prime}}S_{CB}^{\prime} &  =S_{CA}^{^{\prime}}S_{CB}^{\prime}%
=e_{C}^{\mu}\partial_{\mu}\phi_{A}e_{C}^{\nu}\partial_{\nu}\phi_{B}=g^{\mu\nu
}\partial_{\mu}\phi_{A}\partial_{\nu}\phi_{B}\nonumber\\
&  =H_{AB}\equiv\eta_{AB}+\bar{h}_{AB},\label{14a}%
\end{align}
where the symmetry of $S_{AC}^{^{\prime}}$ imposed by constraint $\left(
\ref{7}\right)  $ is used$.$ This shows that the matrix $S_{AB}^{^{\prime}}$
is the square root of the matrix $H_{AB}$ and, hence,%
\begin{align}
S_{AB}+\eta_{AB} &  =\sqrt{\eta_{AB}+\bar{h}_{AB}}\nonumber\\
&  =\eta_{AB}+\frac{1}{2}\bar{h}_{AB}-\frac{1}{8}\bar{h}_{A}^{\,\,\,C}\bar
{h}_{CB}+\frac{1}{16}\bar{h}_{A}^{\,\,\,C}\bar{h}_{C}^{\,\,\,D}\bar{h}%
_{DB}+\cdots.\label{16}%
\end{align}
Substituting this expansion in $\left(  \ref{9}\right)  $ one can easily check
that the obtained Lagrangian will be reduced to $\left(  \ref{3}\right)  $
with Fierz-Pauli quadratic term (In the Appendix we show how this comes out
explicitly by solving perturbatively the equations for $e_{A}^{\mu}$).
Moreover, the action for the scalar fields without higher order terms
$O\left(  S^{3}\right)  ,$ that is,%
\begin{equation}
S_{\phi}=\frac{m_{g}^{2}}{8}\int d^{4}x\,\sqrt{-g}\left[  S^{2}-S_{AB}%
S^{AB}\right]  \label{17}%
\end{equation}
precisely corresponds to the theory \cite{GR1}, which is advocated to be ghost
free\cite{GR1}, \cite{HR}. Now we have rewritten this theory in a form when
the scalar fields comes only to second order. Therefore, if we can show that
the scalar field $\phi^{0}$ becomes dynamical then there is no hope that the
ghost can be canceled in higher orders via field redefinitions. Substituting
\begin{equation}
e_{A}^{\mu}=\delta_{A}^{\mu}+\mathit{l}_{\,\,A}^{\mu},\text{ \qquad}\phi
^{A}=x^{A}+\chi^{A},\label{18}%
\end{equation}
in $\left(  \ref{6}\right)  $ we get%
\begin{equation}
S_{AB}=\mathit{l}_{AB}+\chi_{B,A}+\mathit{l}_{\,\,A}^{\mu}\chi_{B,\mu
}\label{19}%
\end{equation}
in particular,
\begin{equation}
S_{i0}=\mathit{l}_{i0}+\dot{\chi}_{i}+\mathit{l}_{\,\,\,i\,}^{0}\dot{\chi}%
_{0}+\mathit{l}_{\,\,\,i}^{k}\chi_{0,k},\label{20}%
\end{equation}
where dot denotes derivative with respect to time and we raise and lower all
indices with the Minkowski metric. Keeping in action $\left(  \ref{17}\right)
$ only relevant terms and substituting there equation $\left(  \ref{20}%
\right)  $ we find%
\begin{equation}
S_{\phi}=\frac{m_{g}^{2}}{8}\int d^{4}x\,\sqrt{-g}\left[  -S_{i0}%
S^{i0}+...\right]  =\frac{m_{g}^{2}}{8}\int d^{4}x\,\sqrt{-g}\left[
\mathit{l}_{\,\,\,i}^{0}\mathit{l}_{\,\,\,i}^{0}\dot{\chi}_{0}^{2}+...\right]
.\label{21}%
\end{equation}
Notice that the fields $\chi^{A}$ are independent variables and $\mathit{l}%
_{\,\,\,A}^{\mu}$ take care about the constraints. If $\mathit{l}%
_{\,\,\,i}^{0}\neq0$ the field $\chi^{0}$ propagates in a non-trivial
background and there is no hope for this degree of freedom, even in principle,
to be removed by a field redefinition because there are no higher order terms
in the action which could cancel this mode (a perturbative expression for
$\mathit{l}_{\,\,i}^{0}$ is given in the appendix).

Thus we find in a very straightforward manner that the ghost state associated
with $\chi_{0}$ does not propagate except in a non-trivial background with
non-vanishing $\mathit{l}_{i}^{0}.$ However, on this background there is no
way to avoid the ghost. This confirms our previous result \cite{CM2} obtained
through a long calculation, where the equations of motion were analyzed up to
fourth order, and proves that the nonlinear BD ghost is unavoidable in massive
gravity (see also \cite{giastudents}).

It is now possible to see that unlike the old non-Higgs like formulations of
massive gravity, there is no problem here with the unacceptable constraint of
vanishing linearized curvature \cite{boul}. The linearized form of equation
(\ref{Einstein}) is given by%
\begin{align}
R_{\mu\nu}  & =-\frac{m^{2}}{4}\left(  S_{\mu\nu}-\eta_{\mu\nu}S\right)  \\
& =-\frac{m^{2}}{4}\left(  \partial_{\mu}\chi_{\nu}-\eta_{\mu\nu}%
\partial^{\rho}\chi_{\rho}\right)
\end{align}
where $\partial_{\mu}\chi_{\nu}=\partial_{\nu}\chi_{\mu}$ by symmetry of
$S_{\mu\nu}.$ Thus the Bianchi identity implies
\begin{align}
0  & =\partial^{\mu}\left(  R_{\mu\nu}-\frac{1}{2}\eta_{\mu\nu}R\right)  \\
& =-\frac{m^{2}}{4}\left(  \partial^{2}\chi_{\nu}+\frac{1}{2}\partial_{\nu
}\partial^{\rho}\chi_{\rho}\right)  \\
& =-\frac{3m^{2}}{8}\partial^{2}\chi_{\nu}%
\end{align}
which is the vanishing of the Laplacian of the scalar fields $\chi_{\nu}$.
This is to be contrasted with the Pauli-Fierz formulation where the Bianchi
identity implies the unacceptable condition $\partial^{\mu}h_{\mu\nu}%
=\partial_{\nu}h$ which is equivalent to the vanishing of linearized curvature.

We also like to mention that there exists a special combination of higher
order terms as function of $S_{AB}$
\begin{equation}%
{\displaystyle\int}
d^{4}x\sqrt{-g}\epsilon_{MNPQ}\epsilon^{ABCD}S_{A}^{M}S_{B}^{N}S_{C}%
^{P}\left(  \frac{c_{1}}{3!}\ \delta_{D}^{Q}\ +\frac{c_{2}}{4!}S_{D}%
^{Q}\right)
\end{equation}
which could be thought of as generalization of the Fierz-Pauli form. 

\appendix\textbf{Appendix}

In this appendix we will solve the constraints explicitly using perturbation
theory to check the consistency of the consideration above. Due to the
constraints $\left(  \ref{5}\right)  $ and $\left(  \ref{6}\right)  $ the
metric can be written as
\begin{equation}
g^{AB}=\eta^{AB}+h^{AB}=\eta^{AB}+\mathit{l}^{AB}+\mathit{l}^{BA}%
+\mathit{l}_{\,\,\,\,C}^{A}\mathit{l}^{BC},
\end{equation}
while the symmetry of $S_{AB}$ gives%
\[
\left(  \delta_{A}^{\mu}+\mathit{l}_{\,\,\,A}^{\mu}\right)  \left(  \eta_{\mu
B}+\partial_{\mu}\chi_{B}\right)  -\left(  \delta_{B}^{\mu}+\mathit{l}%
_{\,\,\,B}^{\mu}\right)  \left(  \eta_{\mu A}+\partial_{\mu}\chi_{A}\right)
=0,
\]
or equivalently
\begin{equation}
\mathit{l}_{BA}-\mathit{l}_{AB}+\partial_{A}\chi_{B}-\partial_{B}\chi
_{A}+\mathit{l}_{\,\,\,A}^{\mu}\partial_{\mu}\chi_{B}-\mathit{l}%
_{\,\,\,B}^{\mu}\partial_{\mu}\chi_{A}=0.
\end{equation}
We now solve equations for the symmetric and antisymmetric parts of
\begin{equation}
\mathit{l}_{AB}=p_{AB}+q_{AB},
\end{equation}
where $p_{AB}=p_{BA}$ and $q_{AB}=-q_{BA},$ perturbatively. With this purpose
we write%
\begin{align*}
p_{AB}  &  =p_{AB}^{\left(  1\right)  }+p_{AB}^{\left(  2\right)  }%
+p_{AB}^{\left(  3\right)  }+\cdots,\\
q_{AB}  &  =q_{AB}^{\left(  1\right)  }+q_{AB}^{\left(  2\right)  }%
+q_{AB}^{\left(  3\right)  }+\cdots,
\end{align*}
which being substituted in the constraint equations give
\begin{align}
p_{AB}^{\left(  1\right)  }  &  =\frac{1}{2}h_{AB},\\
q_{AB}^{\left(  1\right)  }  &  =\frac{1}{2}\left(  \partial_{A}\chi
_{B}-\partial_{B}\chi_{A}\right)  ,\\
p_{AB}^{\left(  2\right)  }  &  =-\frac{1}{8}\left(  h_{AC}+\partial_{A}%
\chi_{C}-\partial_{C}\chi_{A}\right)  \left(  h_{B}^{\,\,\,C}+\partial_{B}%
\chi^{C}-\partial^{C}\chi_{B}\right)  ,\\
q_{AB}^{\left(  2\right)  }  &  =\frac{1}{4}\left(  h_{CA}+\partial_{C}%
\chi_{A}-\partial_{A}\chi_{C}\right)  \partial^{C}\chi_{B}-\frac{1}{4}\left(
h_{CB}+\partial_{C}\chi_{B}-\partial_{B}\chi_{C}\right)  \partial^{C}\chi_{A},
\end{align}
Combining these relations we find%
\begin{align}
S_{AB}  &  =p_{AB}+\frac{1}{2}\left(  \partial_{A}\chi_{B}+\partial_{B}%
\chi_{A}\right)  +\frac{1}{2}\left(  \mathit{l}_{\,\,\,A}^{\mu}\partial_{\mu
}\chi_{B}+\mathit{l}_{\,\,\,B}^{\mu}\partial_{\mu}\chi_{A}\right) \\
&  =\frac{1}{2}\bar{h}_{AB}-\frac{1}{8}\bar{h}_{A}^{\,\,\,C}\bar{h}%
_{CB}+\cdots
\end{align}
and one can continue this procedure to verify that the formal result $\left(
\ref{16}\right)  $ holds.

Having seen that the fields $e_{A}^{\mu}$ depend on the fields $g^{\mu\nu}$
and $\partial_{\mu}\phi_{A}$ through relations that could be only solved
perturbatively, it is essential to impose the constraints through Lagrange
multipliers to get the correct equations of motion. First the $g^{\mu\nu}$
equation of motion gives
\begin{equation}
R_{\mu\nu}-\frac{1}{2}g_{\mu\nu}R=\tau_{\mu\nu}-\frac{1}{2}g_{\mu\nu}\tau,
\end{equation}
and thus
\begin{equation}
\tau_{\mu\nu}=R_{\mu\nu}.
\end{equation}
The $e_{A}^{\mu}$ equations give%
\begin{equation}
\lambda^{AB}\partial_{\mu}\phi_{B}=R_{\mu\nu}e^{\nu A}+\frac{m^{2}}{4}\left(
S^{AB}\partial_{\mu}\phi_{B}-S\partial_{\mu}\phi^{A}\right)  ,
\end{equation}
which is a set of $16$ equations, ten for the Ricci tensor $R_{\mu\nu}$ and
six for the antisymmetric matrices $\lambda^{AB}.$ The $\phi^{A}$ equations of
motion give%
\begin{equation}
\partial_{\mu}\left(  \sqrt{-g}\lambda^{AB}e_{A}^{\mu}\right)  =\frac{m^{2}%
}{4}\partial_{\mu}\left(  \sqrt{-g}\left(  S^{AB}e_{A}^{\mu}-Se^{\mu
B}\right)  \right)  .
\end{equation}
Assuming that the "vierbein" $\partial_{\mu}\phi_{B}=\eta_{\mu B}%
+\partial_{\mu}\chi_{B}$ has an inverse, we deduce that%
\begin{align}
\lambda^{AB} &  =\frac{1}{2}R_{\mu\nu}\left(  e^{\nu A}\left(  \partial_{\mu
}\phi_{B}\right)  ^{-1}-e^{\nu B}\left(  \partial_{\mu}\phi_{A}\right)
^{-1}\right)  ,\\
-\frac{m^{2}}{4}\left(  S^{AB}-\eta^{AB}S\right)   &  =\frac{1}{2}R_{\mu\nu
}\left(  e^{\nu A}\left(  \partial_{\mu}\phi_{B}\right)  ^{-1}+e^{\nu
B}\left(  \partial_{\mu}\phi_{A}\right)  ^{-1}\right)  .\label{Einstein}%
\end{align}
Notice that because of the symmetry of $R_{\mu\nu}$ the first non-vanishing
contribution to $\lambda^{AB}$ is of order two as can be seen by using a
perturbative expansion of $\left(  \partial_{\mu}\phi_{B}\right)  ^{-1}.$

\begin{acknowledgements}
A. H. C \ is supported in part by the National Science Foundation under Grant
No. Phys-0854779. V.M. is supported by TRR 33 \textquotedblleft The Dark
Universe\textquotedblright\ and the Cluster of Excellence EXC 153
\textquotedblleft Origin and Structure of the Universe.\textquotedblright
\end{acknowledgements}


\begin{thebibliography}{9}                                                                                                %


\bibitem {CM}A. Chamseddine and V. Mukhanov, \textquotedblleft Higgs for
graviton: simple and elegant solution,\textquotedblright\ \emph{JHEP}
\textbf{1008}, 011 (2010)

\bibitem {CM1}L. Alberte, A. Chamseddine and V. Mukhanov, \textquotedblleft
Massive Gravity: Resolving the Puzzles,\textquotedblright\ \emph{JHEP}
\textbf{1012,} 023 (2010).

\bibitem {CM2}L. Alberte, A. Chamseddine and V. Mukhanov, \textquotedblleft
Massive Gravity: Exorcising the Ghost,\textquotedblright\ \emph{JHEP}
\textbf{1104}, 004 (2011).

\bibitem {review}K. Hinterbichler, \textquotedblleft Theoretical Aspects of
Massive Gravity,\textquotedblright\ 139 pp. e-Print: arXiv:1105.3735 [hep-th] (2011).

\bibitem {pauli}M. Fierz and W. Pauli, \textquotedblleft On relativistic wave
equations for particles of arbitrary spin in an electromagnetic
field,\textquotedblright\ \emph{Proc. Roy. Soc. Lond. A} \textbf{173}, 211 (1939).

\bibitem {boul}D. G. Boulware and S. Deser, \textquotedblleft Can gravity have
a finite range?,\textquotedblright\ \emph{Phys. Rev. D} \textbf{6}, 3368 (1972).

\bibitem {GR1}C. de Rham, G. Gabadadze, A. Tolley, \textquotedblleft
Resummation of Massive Gravity,\textquotedblright\ arXiv:1011.1232 [hep-th]
(2010) and references therein.

\bibitem {HR}S. Hassan and R. Rosen, \textquotedblleft Resolving the Ghost
Problem in non-Linear Massive Gravity,\textquotedblright\ arXiv:1106.3344
[hep-th] (2011).

\bibitem {giastudents}S. Folkerts, A. Pritzel, N. Wintergerst, to be published (2011).
\end{thebibliography}
\end{document}